\documentclass{llncs}
\usepackage{hyperref}
\usepackage{amsmath}
\usepackage{amsfonts}
\usepackage{float}
\usepackage{graphicx}
\usepackage{booktabs}   
\usepackage[T1]{fontenc} 

\usepackage{tikz}
\usepackage{enumitem}
\usepackage{multirow} 

\usepackage[frozencache=true,cachedir=minted-cache]{minted}

\setminted{
  numbersep=5pt,
  xleftmargin=12pt,
  breakafter=-/\{,
  breakbefore=\\
}

\usepackage{listings}  

\lstdefinelanguage{SMT-LIB}{
  morekeywords={
    declare-const, assert, check-sat, define-fun, forall, or, and, not, 
    exists, if, then, else, let, in, re, str, seq, Int, String, 
    exists, true, false, "=",
  },
  sensitive=true,
  morecomment=[l]{;}, 
}

\newcommand{\cerl}{\mathcal{R}}
\newcommand*{\aut}{\mathcal{A}}
\newcommand*{\NFA}{\mathcal{A}}
\newcommand*{\autb}{\mathcal{B}}
\newcommand*{\transducer}{\mathfrak{T}}
\newcommand*{\Int}{\mathbb{Z}}
\newcommand*{\myvec}[1]{\overrightarrow{#1}}
\newcommand{\cefaout}{\mathcal{O}}
\newcommand{\Lang}{\mathcal{L}}
\newcommand{\Tran}{\mathcal{T}}
\newcommand{\arrseparator}{\sharp}
\newcommand{\todo}[1]{{\color{orange}\textbf{TODO:} #1 \textbf{:ODOT}\color{black}}}
\newcommand{\commentword}[1]{}

\newcommand{\mysplit}{\mathsf{split}}
\newcommand{\mysplitstr}{\mathsf{splitstr}}

\newcommand{\myjoin}{\mathsf{join}}
\newcommand{\mylen}{\mathsf{len}}
\newcommand{\mystrlen}{\mathsf{strlen}}
\newcommand{\myseqlen}{\mathsf{seqlen}}
\newcommand{\myread}{\mathsf{read}}
\newcommand{\mywrite}{\mathsf{write}}

\newcommand{\mymatchall}{\mathsf{matchAll}}
\newcommand{\mymatchallstr}{\mathsf{matchAllstr}}

\newcommand{\myat}{{\sf nth}}
\newcommand{\myfilter}{{\sf filter}}
\newcommand{\myset}[1]{\{#1\}}
\newcommand{\hide}[1]{ }
\newcommand{\arraylogic}{{\sf SeqStr}}
\newcommand{\slarraylogic}{{\sf SeqStr}_{\sf SL}}
\newcommand{\exstrlogic}{{\sf XStr}}
\newcommand{\slexstrlogic}{{\sf XStr}_{\sf SL}}

\newcommand*{\inttype}{{\sf Int}}
\newcommand*{\strtype}{{\sf Str}}
\newcommand*{\seqtype}{{\sf Seq}}

\newcommand{\svarx}{\mathfrak{x}}
\newcommand{\svary}{\mathfrak{y}}
\newcommand{\svarz}{\mathfrak{z}}

\newcommand{\svaru}{\mathfrak{u}}
\newcommand{\svarv}{\mathfrak{v}}
\newcommand{\svarm}{\mathfrak{m}}
\newcommand{\svarn}{\mathfrak{n}}
\newcommand{\svars}{\mathfrak{s}}
\newcommand{\svart}{\mathfrak{t}}

\newcommand{\seqvara}{\alpha}
\newcommand{\seqvarb}{\beta}
\newcommand{\seqvarg}{\gamma}
\newcommand{\concat}{\cdot}
\newcommand{\eqdef}{\stackrel{\mbox{\begin{tiny}def\end{tiny}}}{=}}

\newcommand{\iop}{\bowtie}
\newcommand{\separator}{\dag}
\newcommand{\enc}{{\sf enc}}
\newcommand{\subseq}{{\sf subseq}}
\newcommand{\elem}{{\sf elem}}

\newcommand{\natnum}{\mathbb{N}}
\newcommand{\intnum}{\mathbb{Z}}

\newcommand{\nfa}{{\sf NFA}}
\newcommand{\nft}{{\sf NFT}}

\newcommand{\ostrichseq} {{\sf OSTRICH^{SEQ}}}
\newcommand{\princessarr} {{\sf Princess^{ARR}}}
\newcommand{\princess}{{\sf Princess}}
\newcommand{\seco}{{\sf SeCo}}
\newcommand{\ostrich} {{\sf OSTRICH}}
\newcommand{\seqbase} {{\sf SEQBASE}}
\newcommand{\seqext} {{\sf SEQEXT}}

\newcommand{\nmatch} {{\sf NotMatch}}
\newcommand{\match} {{\sf Matching}}
\newcommand{\ematch} {{\sf EndMatch}}

\DeclareRobustCommand{\polished}[1]{{\color{black}{#1}}}

\title{Decision Procedure for \\ A Theory of String Sequences 
}

\author{Denghang Hu\inst{1,2} \and 
  Taolue Chen\inst{3} \and
  Philipp R\"{u}mmer\inst{4}\and\\ 
  Fu Song\inst{1,2,5} \and
  Zhilin Wu\inst{1,2} 
  }

\institute{
Key Laboratory of System Software and State Key
  Laboratory of Computer Science, Institute of Software, Chinese Academy
  of Sciences, China 
  \and
University of Chinese Academy of Sciences, China
\and 
Birkbeck, University of London, United Kingdom
\and University of Regensburg, Germany
\and Nanjing Institute of Software Technology, China
}

\begin{document}
\maketitle


\begin{abstract}
The theory of sequences, supported by many SMT solvers, can model program data types including bounded arrays and lists. Sequences are parameterized by the element data type and provide operations such as accessing elements, concatenation, forming sub-sequences and updating elements. Strings and sequences are intimately related; many operations, e.g., matching a string according to a regular expression, 
splitting strings, or joining strings in a sequence, are frequently used in string-manipulating programs. Nevertheless, these operations are typically not directly supported by existing SMT solvers, which instead only consider the generic theory of sequences.  
In this paper, we propose a theory of string sequences and study its
satisfiability. We show that, while it is undecidable in general,  the
decidability can be recovered by restricting to the straight-line
fragment. This is shown by encoding each string sequence as a string,
and each string sequence operation as a corresponding string operation. We provide pre-image computation for the resulting string operations with respect to automata, effectively casting it into the generic OSTRICH string constraint solving framework. 
We implement the new decision procedure as a tool $\ostrichseq$, and carry out experiments on benchmark constraints generated from \polished{real-world JavaScript programs, hand-crafted templates and unit tests}. The experiments confirm the efficacy of our approach. 
\end{abstract}


\section{Introduction} \label{sect:intro}


Many real-world applications, such as web applications and programs processing data of type string, involve a multitude of complex operations that convert between strings and string sequences. Typical examples include $\myjoin$ and $\mysplit$, which are present in 
almost all built-in libraries of modern programming languages.
Reasoning about string sequences, also known as extendable string arrays~\cite{JezLMR23}, is therefore a crucial aspect of program analysis. 

Satisfiability Modulo Theory (SMT~\cite{smt_18}) solving provides an automatic way to verify software systems. 
At the moment, there is no standardized sequence theory in SMT-LIB~\cite{smt_standard}. Instead, in verification tools, often the theory of arrays is used, which is limited and does not provide even basic operations like sequence length. Fortunately, \cite{seq_2012} proposed a basic format for a sequence theory, which has been implemented and extended by SMT solvers such as Z3~\cite{Z3} and cvc5~\cite{BarbosaBBKLMMMN22} with practical operations. In particular, \cite{cvc5_seq} presents a calculus in cvc5 for reasoning about string sequences, which extends the proposal~\cite{seq_2012} with \verb|update| (aka \verb|write|) operations. 
The implementation of sequences in Z3 has been documented in the Z3 guide~\cite{z3_seq_doc} and includes operations such as \verb|map| and \verb|foldleft| (but no \verb|update|).

Although existing SMT solvers are effective at reasoning about sequences in general, to the best of our knowledge, none of them directly support functions that convert between strings and string sequences 
such as $\myjoin$ and $\mysplit$. While it is possible to implement these functions using existing ones for some SMT solvers (e.g., Z3 can use \verb|map| and \verb|foldleft| to implement the $\myjoin$ function), usually the solvers do not provide decision procedures for those theories and fail to solve many instances. 
This limitation poses a significant challenge for program analysis that involves such operations.
%
The purpose of our work is to fill this gap. 


\smallskip
\noindent\emph{Contributions.} We propose a logic (aka constraint language) $\arraylogic$ of string sequences, which extends the existing sequence theory in SMT solvers with dedicated operations of string sequences and strings. In addition to the standard string operations (replace/replaceAll, reverse, finite transducers, regular constraints, 
length, indexOf, substring, etc.), our logic emphasizes the interaction between strings and string sequences through operations such as $\mysplit$, $\myjoin$, $\myfilter$, and $\mymatchall$. For instance, $\mysplit$ splits a string into a sequence of substrings according to a regular expression; $\myjoin$ concatenates all the strings in a sequence to obtain a single string. More details can be found in Section~\ref{sec:logic}. Typically, when string sequences are considered, it is natural to consider the integer data type as well; for instance, one usually needs to refer to the length of a sequence or the index of a specific sequence element. As a result, $\arraylogic$ features three data types: integers, strings, and string sequences. In terms of operations, the logic subsumes, to our best knowledge, most string constraint languages that have been proposed in the literature. 

%

Our main focus is the decision procedure for the satisfiability problem of  $\arraylogic$. 
Because of its expressiveness, it is perhaps not surprising that $\arraylogic$ is undecidable in general. To reinstate decidability, we consider the straight-line fragment, which imposes a syntactic restriction on the constraints written in the logic. 
Straight-line formulas naturally arise when verifying programs using
bounded model checking or  symbolic execution, 
which unroll loops in the programs up to a given
depth and convert programs to static single assignment form (i.e. each variable is defined at most once). For example, a majority of the constraints in the standard Kaluza
benchmarks~\cite{SAHMMS10} 
satisfy this condition. (Cf.\ Section~\ref{sect:example} for a concrete example.)

The general strategy of our decision procedure is to encode each string sequence as a string in such a way that all operations for string sequences considered in $\arraylogic$ can be transformed into string operations, possibly involving integers. As such, the decidability of (straight-line) $\arraylogic$ is reduced to string constraints with the integer data type, which was shown to be decidable \cite{atva2020}. Needless to say, this strategy requires overcoming certain technical challenges, as the string constraints resulting from the reduction typically encompass complex (and sometimes non-standard) string operations, which go well beyond the capacity of state-of-the-art string constraint solvers. 
Recall that for the decidability of string constraints with integers, cost-enriched finite automata (CEFA), a variant of cost-register automata,  
were utilized \cite{atva2020}. 
The crux of our technical contributions is thus to compute the backward images of CEFAs under the fairly complex string operations (cf. Section~\ref{subsect:preimage}), 
so the powerful string constraint solving framework  OSTRICH 
\cite{CCH+18,CHL+19} can be harnessed. 

We implement the decision procedure as a new solver $\ostrichseq$ on top of OSTRICH~\cite{CHL+19} and $\princess$~\cite{princess08}. To evaluate the effectiveness of $\ostrichseq$, we curate two benchmark suites of constraints that 
are randomly generated from templates (only using
the operations directly supported by some existing SMT solvers) and extracted from real-world JavaScript programs and unit tests (involving string sequence operations that are not directly supported by existing SMT solvers), respectively. 
%
On both benchmark suites, the experimental results show that $\ostrichseq$ can solve considerably more constraints
than SOTA string solvers including cvc5, Z3, Z3-noodler~\cite{ChenCHHLS24}, $\ostrich$ and $\princessarr$~\cite{ruemmer2024princess}, while being largely as efficient as most of them. 


\smallskip
\noindent\emph{Organization.} 
Section~\ref{sect:example} presents a motivating example. Section~\ref{sec:prelim} gives the preliminaries. 
Section~\ref{sec:logic} defines the logic $\arraylogic$ of string sequences. Section~\ref{sect:procedure} presents the decision procedure. Section~\ref{sect:exp} presents the benchmarks and
experiments. 
Section~\ref{sect:related} discusses the related work. The paper is concluded in Section~\ref{sect:conc}.

\section{Preliminaries}\label{sec:prelim}


Let $\natnum$ denote the set of natural numbers. For $1\leq n\in \natnum$, let $[n]:=\{1, \ldots, n\}$, and for $m<n \in \natnum$, let $[m,n] := \{j \mid m \le j \le n\}$. We also use standard quantifier-free/existential \emph{linear integer arithmetic} (LIA) formulas, which are typically ranged over by $\phi, \varphi$, etc. For an integer term $t$, we use $t[t_1/t_2]$ to denote the term obtained by replacing $t_2$ with $t_1$ where $t_1$ and $t_2$ are integer terms.

\smallskip
\polished{\noindent \textbf{Strings, languages, and transductions.}} We fix an alphabet $\Sigma$, i.e., a finite set of letters. 
A \emph{string} over $\Sigma$ is a finite sequence of letters from $\Sigma$. We use $\Sigma^*$ to denote the set of strings over $\Sigma$ and $\varepsilon$ to denote the empty string. 
%
A \emph{language} over $\Sigma$ is a subset of $\Sigma^*$. We will use $L_1, L_2, \dots$ to denote languages. For two languages $L_1, L_2\subseteq \Sigma^*$, $L_1 \cup L_2$ denotes the union of $L_1$ and $L_2$, and $L_1 \cdot L_2$  denotes the concatenation of $L_1$ and $L_2$, that is,  $\{u_1 \cdot u_2 \mid u_1 \in L_1, u_2 \in L_2\}$. For a language $L\subseteq\Sigma^*$, we define the complement of $L$ as $\Bar{L} = \{w\in\Sigma^*\mid w\not\in L\}$, moreover, we define $L^n$ for $n \in \natnum$, the \emph{iteration} of $L$ for $n$ times, inductively as: $L^0=\{\varepsilon\}$ and $L^{n} =L \cdot L^{n-1}$ for $n > 0$. We also use $L^*$ to denote the iteration of $L$ for arbitrarily many times, that is, $L^* = \bigcup_{n \in \natnum} L^n$, and let $L^+ = \bigcup_{n>0} L^n$. \polished{A \emph{transduction} over $\Sigma$ is a binary relation over $\Sigma^*$, namely, a subset of $\Sigma^* \times \Sigma^*$.}

Regular expressions over $\Sigma$ are defined in a standard way, i.e., 
$$e \eqdef \emptyset \mid \varepsilon \mid a \mid e + e \mid e \concat e \mid e^*  \text{where}\  a \in \Sigma.$$

The language $\Lang(e)\subseteq\Sigma^*$ of a regular expression $e$ is defined 
	inductively as: $\Lang(\emptyset) =\emptyset$,
	$\Lang(\varepsilon) =\{\varepsilon\}$,
	$\Lang(a)= \{a\}$,
	$\Lang(e_1 + e_2) = \Lang(e_1) \cup \Lang(e_2)$,
	$\Lang(e_1 \concat e_2) = \Lang(e_1) \cdot \Lang(e_2)$,
	$\Lang(e_1^*)=(\Lang(e_1))^*$. \polished{A \emph{regular language} is a language that can be defined by a regular expression.} Moreover, $\Bar{e}$ is the complement of $e$, that is, $\Lang(\Bar{e}) = \{w\mid w\not\in \Lang(e)\}$

\hide{As $+$ is associative and commutative, we write $(e_1 + e_2) + e_3$ as $e_1 + e_2 + e_3$ for brevity. We use the abbreviation $e^+ \equiv e \concat e^*$. Moreover, for $\Gamma = \{a_1, \cdots, a_n\}\subseteq \Sigma$, we may overload $\Gamma$ (resp. $\Gamma^\ast$) as the regular expression $a_1 + \cdots + a_n$ (resp. $(a_1 + \cdots + a_n)^\ast$). 
In addition, $|e|$ denotes the number of symbols occurring in $e$.}

\smallskip
\noindent \textbf{Automata.} A \emph{(nondeterministic) finite automaton} (NFA) $\NFA$ is a 5-tuple $(Q, \Sigma, \delta,$ $ I, F)$, where $Q$ is a finite set of states, $\Sigma$ is a finite alphabet, $\delta \subseteq Q \times \Sigma \times Q$ is the transition relation, $I,F \subseteq Q$ are the sets of initial and final states respectively. For readability, we write a transition $(q, a, q') \in \delta$ as $q \xrightarrow[\delta]{a} q'$ (or simply $q \xrightarrow{a} q'$). 
A \emph{run} of $\NFA$ on a string $w = a_1 \cdots a_n$ is a sequence of transitions $q_0 \xrightarrow{a_1} q_1 \cdots q_{n-1} \xrightarrow{a_n} q_n$ with $q_0 \in I$. The run is \emph{accepting} if $q_n \in F$.
A string $w$ is accepted by an NFA $\NFA$ if there is an accepting run of $\NFA$ on $w$. In particular, the empty string $\varepsilon$ is accepted by $\NFA$ if $I \cap F \neq \emptyset$. The language of $\NFA$, denoted by $\Lang(\NFA)$, is the set of strings accepted by $\NFA$. In addition, an NFA can be built to recognize the language of each given regular expression, and vice versa.

A \emph{(nondeterministic) finite transducer (\nft)} $\transducer$ is an extension of {\nfa} with outputs. Formally, an {\nft} $\transducer$ is a 5-tuple $(Q, \Sigma, \delta, I, F)$, where $Q, \Sigma, I, F$ are the same as in {\nfa} and the transition relation $\delta$ is a finite subset of $Q \times \Sigma \times Q \times \Sigma^*$. For readability, we write a transition $(q, a, q', u) \in \delta$ as $q \xrightarrow[\delta]{a, u} q'$ or $q \xrightarrow{a, u} q'$. A run of $\transducer$ over a string $w=a_1 \cdots a_n$ is a sequence of transitions $q_0 \xrightarrow{a_1, u_1} q_1\xrightarrow{a_2, u_2} q_2\cdots q_{n-1} \xrightarrow{a_n, u_n} q_n$
where $q_0 \in I$. Similar to {\nfa}, the run is  \emph{accepting} if $q_n \in F$. The string $u_1 \cdots u_n$ is called the \emph{output} of the run. The transduction $\Tran(\transducer)\subseteq \Sigma^*\times \Sigma^*$ defined by $\transducer$ is the set of string pairs $(w, u)$ such that there is an accepting run of $\transducer$ on $w$, with the output $u$.
\begin{definition}[Cost-enriched finite automata]
A cost-enriched finite automaton (CEFA for short) $\aut$ is a 6-tuple $(R, Q, \Sigma, \delta, $ $I, F)$ where
	\begin{itemize}[topsep=0pt,partopsep=0pt]
            \item $R = \{r_1, \cdots, r_k\}$ is a finite set of registers,
		\item $Q, I, F$ are the same as in {\nfa}, i.e., the set of states, the set of initial states, and the set of final states, respectively, 
		\item $\delta \subseteq Q \times \Sigma \times Q \times \Int^{k}$ is a transition relation, where $\Int^{k}$ represents the values to update the registers in $R$. 
	\end{itemize}
We write $R_\aut$ for the set of registers of $\aut$
and represent it as a vector $(r_1, \cdots, r_k)$. Accordingly, updates $r_i:= r_i + v_i$ for all $i\in [k]$  are simply identified as the vector $\vec{v}=(v_1, \cdots, v_k)$, i.e., $r_i$ is  incremented by $v_i$ for each $i \in [k]$. Typically, we write a transition $(q, a, q', \vec{v}) \in \delta$ as $q \xrightarrow[\vec{v}]{a} q'$.
\end{definition}    
Intuitively, CEFAs add write-only cost registers to finite automata, where ``write-only'' means that the cost registers can only be written/updated but cannot be read, i.e., they cannot be used in the guards of the transitions.

A \emph{run} of the CEFA $\aut$ on a string $w = a_1 \cdots a_n$ is a sequence of transitions $q_0 \xrightarrow[\myvec{v_1}]{a_1} q_1 \cdots q_{n-1}\xrightarrow[\myvec{v_n}]{a_n} q_n$ such that $q_0 \in I$ and $q_{i-1} \xrightarrow[\myvec{v_i}]{a_i} q_i$ for each $i \in [n]$. 
%
It is \emph{accepting} if $q_n \in F$, and 
the vector $\myvec{c}= \sum_{i \in [n]} \myvec{v_i}$ is defined as the \emph{cost} of the run. 
(Note that all registers are initialized to zero.) 
%
We write $\myvec{c} \in \aut(w)$ if there is an accepting run of $\aut$ on $w$ whose cost is $\myvec{c}$.  The language of a CEFA $\aut$, denoted by $\Lang(\aut)$, is defined as the set of pairs $\{(w, \myvec{c})\in\Sigma^*\times \Int^{|R|} \mid  \myvec{c} \in \aut(w)\}$.
In particular, if $I \cap F \neq \emptyset$, then $(\varepsilon, \myvec{0}) \in \Lang(\aut)$. 
We denote by $\Lang_1(\aut)$ the language $\{w\in\Sigma^*\mid \exists \myvec{c}\in \Int^{|R|}.\ (w, \myvec{c})\in \Lang(\aut)\}$.

\polished{
\smallskip
\noindent \textbf{Pre-images under string functions.} Consider a language $L \subseteq \Sigma^* \times \Int^{k_0}$ defined by a CEFA $\aut =(R, Q, \Sigma, \delta, I, F)$ with the registers $R= (r_1, \cdots, r_{k_0})$ and a function $f: (\Sigma^* \times \Int^{k_1}) \times \cdots \times (\Sigma^* \times \Int^{k_l}) \rightarrow \Sigma^*$.  
For each $i \in [k_0]$, let $r^{i}_1, \cdots, r^{i}_l$ be the freshly introduced registers,
and 
$\vec{t} = (t_1, \cdots ,t_{k_0})$ be a vector of LIA formulas such that $t_i$ is a linear combination of $r^{i}_1, \cdots, r^{i}_l$ for each $i \in [k_0]$. 

The pre-image of $L$ under $f$ with respect to $\vec{t}$, denoted by $f^{-1}_{\vec{t}}(L)$, is 
%
a relation 
\begin{center}
    $\cerl\subseteq  (\Sigma^* \times \Int^{k_1 + k_0}) \times \cdots \times (\Sigma^* \times \Int^{k_l + k_0})$
\end{center}
that comprises tuples of the form  
$((w_1, (\myvec{c_1}, \myvec{d_1})), \cdots, (w_l, (\myvec{c_l}, \myvec{d_l})))$ 
such that
\begin{itemize}
\item 
for every $j\in [l]$,
$\myvec{c_j} \in \Int^{k_j}$ and $\myvec{d_j} = (d^{1}_{j}, \cdots, d^{k_0}_{j}) \in \Int^{k_0}$, 
\item let 
$w_0 = f((w_1, (\myvec{c_1}, \myvec{d_1})), \cdots, (w_l, (\myvec{c_l}, \myvec{d_l})))$ and  
\begin{center}
    $\myvec{d'} = \left(t_1\left[d^{1}_{1}/r^{1}_1, \cdots, d^{1}_{l}/r^{1}_l\right], \cdots, t_{k_0}\left[d^{k_0}_{1}/r^{k_0}_{1}, \cdots, d^{k_0}_{l}/r^{k_0}_{l}\right] \right),$
\end{center}
it holds that 
$(w_0, \myvec{d'}) \in L$.
\end{itemize}
%
%
%
%

The pre-image $f^{-1}_{\vec{t}}(L)$ is \emph{CEFA-definable} if $f^{-1}_{\vec{t}}(L) = \bigcup_{i=1}^n \Lang(\aut_{i,1}) \times \cdots \times \Lang(\aut_{i,l})$ for some $n\ge 1$, where for all $i\in [n]$ and $j\in [l]$, $\aut_{i,j}$ is a CEFA such that $\Lang(\aut_{i,j})\subseteq \Sigma^* \times \Int^{k_j}$.  
Finally, a pre-image of $L$ under $f$, denoted by $f^{-1}(L)$, is a pair $(\cerl,\vec{t})$ such that $\cerl = f^{-1}_{\vec{t}}(L)$.
%

The core concept of \emph{pre-image computation} for the function $f$ involves determining the vector $\vec{t}$ of LIA formulas and, for each CEFA $\aut$, computing $f^{-1}_{\vec{t}}(\Lang(\aut))$ represented as a finite set of CEFAs. 
Intuitively, we remove the string equalities of the form $y = f(x_1, \myvec{it_1}, \cdots, x_l, \myvec{it_l})$ (where $\myvec{it_1}$, $\cdots$, $\myvec{it_l}$ are integer expressions) one by one by computing the pre-images and adding the resulting CEFA membership constraints for $x_1, \cdots, x_l$. In the end,  the original string constraint is transformed into a conjunction of CEFA membership constraints and LIA formulas, whose satisfiability checking 
is known to be PSPACE-complete~\cite{atva2020}.
}

%





\section{Motivating Example} \label{sect:example}


To motivate and illustrate our approach, we consider a \polished{JavaScript} snippet shown in Listing~\ref{lst:js_example}, which is extracted, with a slight adaptation, from a real-world example.\footnote{\url{https://github.com/hgoebl/mobile-detect.js/blob/master/mobile-detect.js}}
Here, the function \texttt{prepareVersionNo} extracts from the input string \texttt{version} a version number in the floating-point number format. 
Specifically, it splits  \texttt{version}  into a sequence/array {\tt numbers} according to the non-numeric characters. If {\tt numbers} contains exactly one element, {\tt numbers[0]} is returned; otherwise, the function returns the concatenation of {\tt numbers[0]}, the dot, and the string obtained by concatenating the other elements of {\tt numbers}.  For instance, if {\tt version = 12a56b23}, the output 
is {\tt 12.5623}.

To verify the functional correctness of  \texttt{prepareVersionNo}, we specify the following precondition and postcondition.  
\begin{itemize}[topsep=0pt,partopsep=0pt]
\item The precondition {\tt version $\in$ \verb|[0-9]+([0-9a-zA-Z._ /-])*|} requires that the input string starts with a decimal number, followed by a string of digits, letters, dot \verb|.|, underline  \verb|_|, blank symbol, slash \verb|/|, or hyphen \verb|-|.
\item The postcondition {\tt result} $\in$ \verb|[0-9]+(\.[0-9]*)?| ensures that the output string starts with a decimal number, possibly followed by a dot \verb|.| and a decimal number. 
\end{itemize}
 
\begin{lstlisting}[float=t,caption={JavaScript code snippet: The motivating example}, label = {lst:js_example}]
  function prepareVersionNo(version) {
    // precondition: version in [0-9]+([0-9a-zA-Z._ /-])*
    numbers = version.split(/[a-z._ \/\-]/i);
    if (numbers.length === 1) { // path-1
      result = numbers[0];     }
    if (numbers.length > 1) { // path-2
      temp = numbers[0] + ".";
      numbers1 = numbers.slice(1, numbers.length);
      result = temp + numbers1.join("");      }
    return Number(result);
    // postcondition: result in [0-9]+(\.[0-9]*)?
  }
\end{lstlisting}

\begin{lstlisting}[float=t,language=SMT-LIB, caption={SMT formula for path-2 in {\tt prepareVersionNo}}, label={lst:js_example_constraints},belowskip=-2\baselineskip,aboveskip=-0.8\baselineskip]
; precondition 
(assert (str.in_re version preReg)) 
(assert (= numbers (str.splitre version splitReg)))
(assert (< 1 (seq.len numbers)))
(assert (= temp (str.++ (seq.nth numbers 0) ".")))
(assert (= numbers1
           (seq.extract numbers 1 (- (seq.len numbers) 1))))
(assert (= result (str.++ temp (seq.join numbers1 ""))))
; postcondition
(assert (not (str.in_re result postReg))) 
\end{lstlisting} 

We apply symbolic execution to verify \texttt{prepareVersionNo}. 
We enumerate its execution paths and check for each path that, provided the input string satisfies the precondition, the output string satisfies the postcondition after the executing the path. 
There are three execution paths: path-1 taking \verb|numbers.length===1|, path-2 taking \verb|numbers.length>1|, and path-3 where both conditions are false.
As an example, we consider path-2, and 
the symbolic execution 
reduces to deciding the satisfiability of the SMT formula in Listing~\ref{lst:js_example_constraints}, where {\tt preReg} and {\tt postReg} denote the regular expressions in the precondition and postcondition, i.e.,
 (\verb|[0-9]+([0-9a-zA-Z._ /-])*|) and (\verb|[0-9]+(\.[0-9]*)?|), respectively, and {\tt splitReg} represents the regular expression \verb|[a-zA-Z._ /-]|. 

This SMT formula involves various operations of string sequences and strings, including sequence length {\tt seq.len}, string split {\tt str.splitre}, sequence read {\tt seq.nth}, string concatenation {\tt str.++},
sequence extract {\tt seq.extract}, sequence join {\tt seq.join}, as well as regular constraints  {\tt str.in\_re}. 
While state-of-the-art SMT solvers such as Z3 and cvc5 can solve constraints in the generic sequence theory, they do not directly support complex operations that feature mutual transformations between strings and string sequences, such as {\tt str.splitre} and {\tt seq.join}. 
This motivates us to define an SMT theory of string sequences ($\arraylogic$, cf. Section~\ref{sec:logic}) and investigate its decision procedure (Section~\ref{sect:procedure}). 

\polished{
Due to the undecidability of $\arraylogic$, we focus on 
its straight-line fragment (cf.\ Definition~\ref{def:straight_line}), into which the formula in Listing~\ref{lst:js_example_constraints} falls.  
We encode string sequences as strings, and, accordingly, 
string sequence operations become string operations (Section~\ref{subsect:seq2string}). 
Then we transform the formula in Listing~\ref{lst:js_example_constraints} into the following string constraint:
\begin{gather}
    {\tt version} \in {\tt preReg}\wedge {\tt snumbers} = \mysplitstr_{\tt splitReg}({\tt version})\wedge \nonumber\\ 
    1<\myseqlen({\tt snumbers})-1 \wedge  {\tt temp} = \elem({\tt snumbers}, 0)\cdot \text{'}.\text{'} \wedge  \nonumber\\
    {\tt snumbers1} = \subseq[{\tt snumbers}, 2, \myseqlen({\tt snumbers}) - 2]\ \wedge \label{forml:motivate} \\ 
    {\tt result} = {\tt temp} \cdot \myjoin_{\epsilon}({\tt snumbers1})\wedge {\tt result} \not\in {\tt posReg}. \nonumber
\end{gather}
where ${\tt snumbers}, {\tt snumbers1}$  are string variables that encode the string sequences ${\tt numbers}, {\tt numbers1}$ in Listing~\ref{lst:js_example_constraints}.
%
Note that the resulting string constraint contains new string operations $\tt splitstr, seqlen, elem, \cdots$ introduced in this paper (cf.\ Section~\ref{subsect:seq2string}).}
Subsequently, we employ the decision procedure outlined in \cite{atva2020}, which 
\emph{back propagates} the CEFA membership constraints by computing the pre-images under string functions (see Section~\ref{subsect:preimage}).

In the sequel, we show how to solve the formula in (\ref{forml:motivate}). 
\polished{
\begin{enumerate}[topsep=0pt,partopsep=0pt]
    \item We back propagate the regular membership constraint ${\tt result} \in \overline{{\tt posReg}}$ (where $\overline{\tt posReg}$ denotes the complement of ${\tt posReg}$) by computing the pre-images under the concatenation $\cdot$ and $\myjoin_\epsilon$, and adding the CEFA constraints for $\tt temp$ and $\tt snumbers1$, say, ${\tt temp} \in \aut_1 \wedge {\tt snumbers1} \in \aut_2$ (where ${\tt temp} \in \aut_1$ means that the value of ${\tt temp}$ belongs to $\Lang_1(\aut_1)$, similarly for ${\tt snumbers1} \in \aut_2$). Note that nondeterministic choices may be made here since the pre-images of $\Lang(\overline{{\tt posReg}})$ under $\cdot$ 
     may be a union of products of CEFAs. After the propagation, the string equality ${\tt result} = {\tt temp} \cdot \myjoin_{\epsilon}({\tt snumbers1})$ is removed. 
    %
    %
    \item Then we back propagate the CEFA membership constraint ${\tt temp} \in \aut_1$ for the equality ${\tt temp} = \elem({\tt snumbers}, 0) \cdot \text{'}.\text{'}$ by computing the pre-image of $\Lang(\aut_1)$ under $\cdot$ as well as $\elem$ and adding some CEFA membership constraint for $\tt snumbers$, say, ${\tt snumbers} \in \aut_3$. Similarly, we back propagate the CEFA membership constraint ${\tt snumbers1} \in \Lang(\aut_2)$ for the equality ${\tt snumbers1} =\subseq[{\tt snumbers}, 2,\myseqlen({\tt snumbers})-1]$ by computing the pre-image of $\Lang(\aut_2)$ under $\subseq$ and adding some CEFA membership constraint for $\tt snumbers$, say, ${\tt snumbers} \in \aut_4$. After these back-propagation operations, the two string equalities are removed. 
    \item Finally, we back propagate the CEFA membership  constraint ${\tt snumbers} \in \aut_3 \cap \aut_4$ for ${\tt snumbers} = \mysplitstr_{\tt splitReg}({\tt version})$
 by computing the pre-image of $\Lang(\aut_3 \cap \aut_4$) under the function $\mysplitstr_{\tt splitReg}$ and adding a CEFA membership constraint ${\tt version} \in \aut_5$. Moreover, the equality ${\tt snumbers} = \mysplitstr_{\tt splitReg}({\tt version})$ is removed. 
 
    
    \item In the end, we obtain a string constraint that contains no string equalities, namely, a Boolean combination of CEFA membership constraints and LIA formulas, which are solvable using existing methods.   
\end{enumerate}
Note that here we prioritize the illustration of the main idea over the precision of the description;  
the technical details of the decision procedure are deferred to later sections. 
}

\hide{In the example, the resulting constraint  
contains a complex combination of  {\tt str.replace\_re\_all}, {\tt str.substr} and {\tt str.len} and cannot be solved by Z3 or cvc5. (Z3 returns {\tt unknown} as it does not support the {\tt str.replace\_re\_all} function well; cvc5 fails to solve the problem in 300 seconds.) }

Overall, the formula in Listing~\ref{lst:js_example_constraints} can be transformed into a string constraint containing a complex combination of {\tt str.replace\_re\_all}, {\tt str.substr} and {\tt str.len}, which cannot be solved by Z3 or cvc5 in a reasonable time limit (Z3 returns unknown as it does not support the {\tt str.replace\_re\_all} function; cvc5 fails to solve the stirng constraint in 20 hours).
In contrast, our new solver $\ostrichseq$ solves it in less than 1~second. 

\section{A Theory of String Sequences ($\arraylogic$)}\label{sec:logic}


A \emph{sequence} over the set $X$ is $(a_0, \ldots, a_{n-1})$, where $a_i \in X$ for each $i \in [0,n-1]$, and $n$ is the \emph{length} of the sequence.
We shall focus on \emph{string sequences}, namely, sequences over $\Sigma^*$. We use $()$ to denote the empty sequence.

\smallskip
\noindent\emph{Operations on string sequences and strings.} 
We assume two string sequences $s_1 = (u_0, \ldots, u_{m-1})$ and  $s_2 = (v_0, \ldots, v_{n-1})$.
\begin{itemize}
	\item $s_1 \concat s_2$ is the concatenation of $s_1$ and $s_2$, i.e., $(u_0, \ldots, u_{m-1}, v_0, \ldots, v_{n-1})$.

	      %
	\item $\myat(s_1, i)$ is $u_i$ if $i\in [0,m-1]$, and  undefined otherwise.
	\item $\myjoin_v(s_1)$ joins the elements in $s_1$ with $v$ as the intermediate string, i.e., \polished{$\myjoin_v(s_1) = u_0 v u_1 v u_2 v \cdots v u_{m-1}$}. 
	\item $s_1[i \rightarrow u]$ replaces $u_i$ in $s_1$ by $u$ if $i\in [0,m-1]$,  and is $s_1$ otherwise,  i.e.
	\begin{center}
	 $[s_1[i \rightarrow u] = 
	\left \{
	\begin{array}{l l}
	(u_0, \ldots, u_{i-1}, u, u_{i+1}, \ldots, u_{m-1}), & \mbox{ if } i\in [0,m-1]; \\
	s_1, & \mbox{ otherwise}.
	\end{array}
	\right.
	$    
	\end{center}
	\item $\myfilter_e(s_1)$ filters out all elements in $s_1$ that do not match the regular expression $e$, i.e., $\myfilter_e(s_1) = (u_{i_1}, \ldots, u_{i_k})$  such that $0\le i_1 < \cdots < i_k\le m-1$ and  for all $j \in [0,m-1]$, $u_j \in \Lang(e)$ iff $j \in \{i_1, \ldots, i_k\}$. 
	      In particular, if none of $u_0, \ldots, u_{m-1}$ are in $\Lang(e)$, then $\myfilter_e(s_1) = ()$. 
	      %
	\item $\mysplit_e(u)$ splits a string $u$ into a sequence of substrings according to the regular expression $e$, i.e.,  $\mysplit_e(u) = (u'_1, \ldots, u'_k)$, where $u = u'_1 v_1 u'_2 v_2 \cdots u'_{k-1}v_{k-1} u'_k$ such that (1) for each $i \in [k-1]$, $v_i$ is \polished{the leftmost and longest substring of $u'_i v_i u'_{i+1}\cdots v_{k-1}u'_k$} that matches $e$, and (2) $u'_k$ does not contain any substring that matches $e$.
	      In particular, if $u$ does not contain any substring that matches $e$, then $\mysplit_e(u) = (u)$.

	\item $s_1[i, j]$ is defined as follows: if $i\in [0,m-1]$ and $j \ge 1$, then $s_1[i, j]$ is  the subsequence $(u_i, \ldots, u_{\min(i+j-1,m-1)})$; if $i\in [0,m-1]$ and $j = 0$, then $s_1[i, j]$ is the empty sequence $()$;  if $i\not\in [0,m-1]$ or $j < 0$, then $s_1[i, j]$ is undefined. 

	      %
	\item $\mymatchall_e(u)$ extracts the substrings of $u$ that match the regular expression $e$, 
	 i.e., $\mymatchall_e(u) = (v_1, \ldots, v_{k-1})$ such that $u = u'_1 v_1 u'_2 v_2 \cdots u'_{k-1}v_{k-1} u'_k$ and (1) for each $i \in [k-1]$, $v_i$ is \polished{the leftmost and longest substring of $u'_i v_i u'_{i+1}\cdots v_{k-1}u'_k$}, and (2) $u'_k$ has no substring that matches $e$.

\end{itemize}

\polished{Note that we choose the longest match semantics in $\mysplit_e(u)$ and $\mymatchall_e(u)$
just for illustrating our approach, where
the shortest match semantics can be captured by adapting the automata construction in the pre-image computation.}

\smallskip
\noindent\emph{The logic $\arraylogic$ of string sequences.} $\arraylogic$   
has three data types: $\inttype$ (integer), $\strtype$ (strings), and $\seqtype$ (sequences).
%
We use $u, v, \ldots$ (resp. $\svaru, \svarv, \ldots$) to denote string constants (resp. variables); $m,n, \ldots$ (resp. $\svarm, \svarn, \ldots$) to denote integer constants (resp. variables);
$s, t, \ldots$ (resp. $\svars,\svart, \ldots$) to denote sequence constants (resp. variables); and $e$ to denote regular expressions.

The syntax of $\arraylogic$ is defined as follows:
\[
\begin{tabular}{c l l l r}
	$~$ &	$it$      & $ \eqdef $ & $m \mid \svarm \mid it+it \mid it-it \mid \mystrlen(strt)\mid \myseqlen(seqt) $                                                      & integer terms  \\
		&	$strt$    & $ \eqdef $ & $u \mid \svaru \mid strt \concat strt \mid \myat(seqt, it) \mid \myjoin_u(seqt)    $                      & string terms   \\
		&	$seqt$    & $ \eqdef $ & $(strt) \mid s \mid \svars \mid seqt \concat seqt \mid seqt[it \rightarrow strt] \mid  \myfilter_e(seqt) \mid  $ &                \\
			&          &            & $\ \ seqt[it, it]  \mid \mysplit_e(strt) \mid \mymatchall_e(strt) $                              & sequence terms \\
		&	$\varphi$ & $ \eqdef $ & $ it \iop it  \mid seqt = seqt \mid strt = strt \mid strt \in e  \mid \varphi \wedge \varphi $                   & formulas
	\end{tabular}
    \]
where  $\iop\, \in \{=, \neq, <, \le, >, \ge\}$.
%

Here, $it$, $seqt$ and $strt$ denote integer, sequence and string terms, respectively. In particular,
the integer terms are linear arithmetic expressions where $\myseqlen(\svars)$ (resp. $\mystrlen(\svaru)$) denotes the length of a sequence $\svars$ (resp. string $\svaru$);
the sequence terms are constructed from a singleton sequence, 
sequence constants and sequence variables by applying the aforementioned sequence/string operations.
%
A $\arraylogic$ formula is a conjunction of atomic formulas, each of which is of the form $it_1 \iop it_2$ where $\iop\, \in \{=, \neq, <, \le, >, \ge\}$,  a sequence equality $seqt_1 = seqt_2$, or a string equality $strt_1 = strt_2$.
%
%
Note that sequence and string inequalities can be expressed as well. 
For instance (using disjunctions and quantifiers for the sake of presentation):
%
 \begin{align*}
    \svarx \neq \svary & \equiv& \exists \svarx_1, \svarx_2, \svary_2. \bigvee \limits_{a, b \in \Sigma, a \neq b} (\svarx = \svarx_1 a \svarx_2 \wedge \svary = \svarx_1 b \svary_2) \vee \exists \svarz. \bigvee_{a \in \Sigma} (\svarx = \svary a \svarz \vee \svary = \svarx a \svarz) \\
		\seqvara \neq \seqvarb &\equiv & (\exists \seqvara_1, \seqvara_2, \seqvarb_2, \svarz_1, \svarz_2.\ \seqvara = \seqvara_1 \concat (\svarz_1) \concat \seqvara_2 \wedge \seqvarb = \seqvara_1 \concat (\svarz_2) \concat \seqvarb_2 \wedge \svarz_1 \neq \svarz_2)\ \vee \\
		                   &   & 
                               \exists \seqvarg, \svarz.\ \seqvara  = \seqvarb \concat (\svarz) \concat \seqvarg \vee    \seqvarb  = \seqvara \concat (\svarz) \concat \seqvarg.
\end{align*}
\polished{Though rewriting inequalities produces disjunctions, they can still be handled in the DPLL(T) framework~\cite {dpll_t}. Hence, we focus on the disjunction-free fragment.}


\begin{proposition}\label{prop-undec}
	The satisfiability of $\arraylogic$ is undecidable.
\end{proposition}
%

We define the \emph{straight-line fragment} $\slarraylogic$ of $\arraylogic$. It is easy to see that by introducing fresh 
variables, we can rewrite a $\arraylogic$ formula to a form in which the left-hand side of each equality $seqt_1 = seqt_2$ (resp. $strt_1 = strt_2$) is a sequence (resp. string) variable.
%
Note that the original and rewritten formulas are equisatisfiable. 
Henceforth, we assume that all sequence/string equalities satisfy this constraint unless otherwise stated. 


\begin{definition}[Straight-line fragment]\label{def:straight_line}
	Given a $\arraylogic$ formula $\varphi$, 
	 the \emph{dependency graph} $G_\varphi$ of  $\varphi$
	 is a directed graph $(V,E)$ such that
	$V$ is  the set of string and sequence variables in $\varphi$, and 
	 $(v_1, v_2)\in E$ 
	iff  $v_1 = {\tt rhs}$ is a sequence or string equality in $\varphi$ and $v_2$ occurs in ${\tt rhs}$ except for integer terms.
	%

$\varphi$ is \emph{straight-line} if $\varphi$ is disjunction-free, and each string/sequence variable occurs as the left-hand side of equalities at most once; moreover, $G_\varphi$ is acyclic.

\end{definition}

\begin{example}
$\svars = \mysplit_e(\svaru) \wedge \svart = \svars[1, \myseqlen(\svars) - 1]$ is  straight-line, while  $\svars = \svars[1, \myseqlen(\svars) - 1]$ is not, since there is a self-dependency on $\svars$.
\end{example}



\hide{
	\begin{definition}[Relational constraints ]
		Relational constraints $\phi$ are defined by the following rules:
		\begin{align*}
			\phi \ ::= \  & A = B[m:n] \mid A = \transducer(B) \mid x=\myjoin(A,u) \mid A = \mysplit(x,e) \mid                       \\
			              & x=\myread(A, m) \mid A = \mywrite(B, x ,m) \mid x = \transducer(y) \mid x=y\cdot z \mid \phi \wedge \phi
		\end{align*}
	\end{definition}
	\begin{definition}[Straight-line relational constraints]
		A relational constraint $\phi$ is straight-line if $\phi = \bigwedge\limits_{1\leq i \leq m} \chi_i = P_i$ such that
		\begin{itemize}
			\item $\chi_1,\cdots,\chi_m $ are mutally distinct string varibles and array varibles.
			\item For each $i\in [m] $, all the varibles in $P_i$ are in the set $\{\chi_1,\cdots,\chi_{m-1} \}$.
		\end{itemize}
	\end{definition}
	\begin{definition}[Straight-line formula]
		A formula $F$ of $\arraylogic$ is straight-line iff all relational constraints in $F$ are straight-line.
	\end{definition}
}

\section{The Decision Procedure} \label{sect:procedure}


\begin{theorem}\label{thm-main}
	The satisfiability of $\slarraylogic$ is decidable.
\end{theorem}

The general idea of the decision procedure is to encode each string sequence as a string, based on which all string sequence operations in $\slarraylogic$ can be transformed into string operations. 
It utilizes the framework~\cite{atva2020} for solving straight-line string constraints with integer data type,  \polished{as illustrated in Section~\ref{sect:example}}. There are, however, certain technical challenges. For instance, 
$seqt[it \rightarrow strt]$ and $seqt[it, it]$ cannot be captured directly by standard string operations,  
so new ones have to be provided whose pre-images need to be computed. 
%

\subsection{From string sequences to strings}\label{subsect:seq2string}


We fix a symbol $\separator\notin\Sigma$ and let $\Sigma_{\separator}:=\Sigma \cup \{\separator\}$.
Each string sequence  $s = (u_1, \cdots, u_m)$ over $\Sigma$ is encoded as a string  \polished{$\enc(s):=\separator u_1\separator u_2 \separator \cdots \separator u_m \separator$} over $\Sigma_{\separator}$. 
Note that the strings 
resulted from the encoding should match the regular expression $\separator (\Sigma^* \separator)^*$ and the string $\separator$ encodes the empty sequence.

With this encoding of string sequences as strings, each string sequence operation is also transformed into the corresponding \emph{string operation}. 
%
\begin{description}
\item[sequence concatenation $s_1 \concat s_2$:] 
string concatenation  $\enc(s_1) \concat \enc(s_2)$. 
\item[sequence  write {$s[i \rightarrow u]$}: ] 
$\mywrite(\enc(s), i+1, u)$, which replaces the substring of $\enc(s)$ between the $(i+1)$-th occurrence of $\separator$ and the $(i+2)$-th occurrence of $\separator$, by $u$, if $i\in[0,n-2]$, where $n$ is the number of occurrences of $\separator$ in $\enc(s)$, and is undefined otherwise (i.e. $i\not\in[0,n-2]$). 
\hide{Note that this operation is not a typical string replace operation.  (Recall that the indices of elements in sequences start from $0$ and $n$, i.e. the number of occurrences of $\separator$ in $\enc(s)$, is one more than the length of $s$.) }

\item[sequence filter operation $\myfilter_e(s)$:] 
$\myfilter_e(\enc(s))$, which removes every substring of $\enc(s)$ between two consecutive occurrences of $\separator$ that does not match the regular expression $e$. (For instance, if $e = a^* b$ and $s = (ab, ac, aab)$, then $\myfilter_e(\enc(s)) = \separator ab \separator aab \separator$.)
\item[subsequence operation {$s[i, j]$}:] 
$\subseq(\enc(s), i+1, j)$, which keeps only the substring of $\enc(s)$ between the $(i+1)$-th occurrence and the $\min(i+j, n-1)$-th occurrence of $\separator$, if $i\in[0,n-2]$, where $n$ is the number of occurrences of $\separator$ in $\enc(s)$, and  is undefined otherwise (i.e. $i\not\in[0,n-2]$). 
\item [$\mysplit_e(u)$:] string operation $\mysplitstr_e(u)$ that transforms $u = u'_1 v_1 u'_2 \cdots u'_{k-1} v_{k-1} u'_k$ into  $\separator u'_1 \separator u'_2 \cdots u'_{k-1} \separator u'_k \separator$, where for each $i \in [k-1]$,  \polished{$v_i$} is the leftmost and longest matching of $e$ in \polished{$u'_i v_i u'_{i+1}\cdots v_{k-1}u'_k$}, and $u'_k$ does not contain any substring that matches $e$. 
%
%
\item[$\mymatchall_e(u)$:] 
$\mymatchallstr_e(u)$ that transforms $u = u'_1 v_1 u'_2 v_2 \cdots u'_{k-1}v_{k-1} u'_k$ into 
$\separator v_1 \separator \cdots \separator v_{k-1} \separator$, where for each $i \in [k-1]$, $v_i$ is the leftmost and longest occurrence of $e$ in \polished{$u'_i v_i u'_{i+1}\cdots v_{k-1}u'_k$}, moreover, $u'_k$ does not contain any substring that matches $e$.

%
\item[sequence read operation $\myat(s, i+1)$:] $\elem(\enc(s), i+1)$ that keeps only the substring between the $(i+1)$-th occurrence  and the $(i+2)$-th occurrence of $\separator$, if $i\in[0,n-2]$, where $n$ is the number of occurrences of $\separator$ in $\enc(s)$, and is undefined otherwise (i.e. $i\not\in[0,n-2]$). (For instance, if $s = (ab, ac)$ and $i = 0$, then $\elem(\enc(s), 1) = \elem(\separator ab \separator ac \separator, 1) = ab$.)
\item[sequence join operation $\myjoin_u(s)$:] $\myjoin_u(\enc(s))$ that removes the first and the last occurrences of $\separator$ and replaces all the remaining occurrences of $\separator$ by $u$.
\item[sequence length operation $\myseqlen(s)$:] $\myseqlen(\enc(s))$ - 1 where $\myseqlen(\enc(s))$ counts the number of occurrences of $\separator$ in $\enc(s)$.
\end{description}

Hence, we obtain atomic string constraints 
of the following forms:
\[\begin{array} {l}
\svarx = u \mid \svarx = \svary \mid \svarx = \svary \concat \svarz \mid \svarx = \transducer(\svary) \mid \svarz = \mywrite(\svarx, it, \svary) \mid  \svary  = \myfilter_e(\svarx)  \mid   \\
\svary = \mysplitstr_e(\svarx) \mid \svary = \subseq(\svarx, it_1, it_2) \mid   
 \svary = \mymatchallstr_e(\svarx) \mid \svary = \elem(\svarx, it)  \mid \\
 \svary = \myjoin_u(\svarx) \mid it = \mystrlen(\svarx) \mid it = \myseqlen(\svarx),
\end{array}\]
where $\transducer$ is a finite-state transducer. 

We denote by $\exstrlogic$ 
the class of string constraints which are 
positive Boolean combinations (no negation) of the atomic string constraints, and  
$\slexstrlogic$ denotes its straight-line fragment.  
For each formula $\varphi$ in $\slarraylogic$, $\enc(\varphi)$ is the resulting formula in $\exstrlogic$. We have the following result.

\begin{proposition} 
For each  constraint $\varphi$ in $\arraylogic$, 
	$\varphi$ and $\enc(\varphi)$ are equisatisfiable.
	Moreover, if $\varphi$ is in $\slarraylogic$, then $\enc(\varphi)$ is in $\slexstrlogic$.
\end{proposition}
\begin{example}
Consider the constraint $\varphi:= (\svars_1 = \svars_0\cdot(u,v)\wedge\myseqlen(\svars_1) < 2)$ in  $\slarraylogic$ where $\svars_0$ is a string variable and $u,v$ are string constants. 
Then, its string encoding is $\enc(\varphi):= (\enc(\svars_1) = \enc(\svars_0)\cdot \separator u \separator v \separator \wedge \myseqlen(\enc(\svars_1)) - 1  < 2)$, which is unsatisfiable, as there are at least three occurrences of $\separator$ in $\enc(\svars_1)$. \qed
\end{example}

The satisfiability of $\slarraylogic$ is now reduced to that of $\slexstrlogic$. 
The operations $\myfilter_e$, $\mysplitstr_e$, $\mymatchallstr_e$ and $\myjoin_u$ can all be represented by a finite-state transducer (see Appendix \ref{sect:appdenix}), 
so the results in~\cite{atva2020} are directly applicable. 
In the next section, we show that the pre-images under the remaining string operations, i.e., 
$\mywrite$, $\subseq$, $\elem$ and $\myseqlen$,  can be effectively computed.

\subsection{Computing the pre-images under $\mywrite$, $\subseq$, $\elem$ and $\myseqlen$} \label{subsect:preimage}

We now show how to compute the pre-images under $\mywrite$, $\subseq$, $\elem$ and $\myseqlen$. Note that the pre-image computation utilizes an NFA $\aut_0$ (resp. $\aut_1$) to format the strings that represent the string sequences (resp. strings), namely, $\aut_0$ (resp. $\aut_1$) recognizes the language $\separator (\Sigma \separator)^*$ (resp. $\Sigma^*$). 

\paragraph*{Pre-image under $\mywrite$.}
Let $\aut = (R, Q, \Sigma_{\separator}, \delta, I, F)$ be a CEFA.  $\mywrite^{-1}(\Lang(\aut))$ is computed as the collection $((\autb_{(p, q)} \cap \aut_0, \aut_{R_2/R}[p, q] \cap \aut_1), t_{(p,q)})_{(p, q) \in Q \times Q}$, where  $\autb_{(p, q)}$, $ \aut_{R_2/R}[p, q]$ and $t_{(p,q)}$ are constructed as follows:
\begin{itemize}
\item $\autb_{(p, q)}$ is constructed by the following procedure: 
\begin{itemize}
%
\item We create two copies of $R$, say $R_1, R_2$. Moreover, we introduce a fresh cost register, say $r'$, to count the number of occurrences of $\separator$. 
\item We run $\aut$ on the input string and increase the cost register $r'$ each time when $\separator$ is read. Moreover, the registers in $R_1$, instead of $R$, are updated in the transitions. 
\item When the current symbol is $\separator$, we nondeterministically choose to stop increasing $r'$ and pause the running of $\aut$ at the state $p$. Let us call this position as the pause position.
\item Finally, when reading the first $\separator$ after the pause position, we resume the run of $\aut$ at $q$, where the registers in $R_1$ (but not $r'$) are updated. 
\end{itemize}

\item $\aut_{R_2/R}[p, q]$ is obtained from $\aut[p, q]$ by replacing each register in $R$ with the corresponding copy in $R_2$, where $\aut[p, q]$ is the sub-automaton of $\aut$ that accepts the strings starting from the state $p$ and ending at the state $q$.
\item $t_{(p, q)} := (t_{(p, q), r})_{r \in R}$ and 
$t_{(p, q), r} = r^{(1)} + r^{(2)}$ \polished{where $r^{(1)}$ and $r^{(2)}$ are the copies of $r$} for each $r \in R$.
\end{itemize}
Note that in the backward propagation of $\aut$ with respect to an equality $\svarz = \mywrite(\svarx, it, \svary)$, the constraint $r' = it $ is added to the LIA constraint to assert that $r'$ is equal to the index $it$. 
Formally, $\mywrite^{-1}(\Lang(\aut))$ is a finite collection of $(\autb_{(p,q)}  \cap \aut_0, \aut_{R_2/R}[p, q] \cap \aut_1, t_{(p,q)})$, where $(p, q) \in Q \times Q$, and $\autb_{(p,q)} = (R_1 \cup \{r'\}, Q', \Sigma_{\separator}, \delta', I', F')$ such that  $Q' = Q \times \{pre, idle, post\}$, $I'  = I \times \{pre\}$, $F'  = F \times \{post\}$, and $\delta'$ comprises:

\begin{itemize}
\item $((q_1, pre), a, (q_2, pre), (\myvec{v}, 0))$ such that $(q_1, a, q_2, \myvec{v}) \in \delta$ and $a \in \Sigma$ (thus $a \neq \separator$), 
\item $((q_1, pre), \separator, (q_2, pre), (\myvec{v}, 1))$ such that $(q_1, \separator, q_2, \myvec{v}) \in \delta$, 
\item $((q_1, pre), \separator, (p, idle), (\myvec{v}, 1))$ such that $(q_1, \separator, p, \myvec{v}) \in \delta$,
%
%
%
\item $((p, idle), a, (p, idle), (\myvec{0}, 0))$ such that $a \in \Sigma$, 
\item $((p, idle), \separator, (q_2, post), (\myvec{v}, 0))$ such that $(q, \separator, q_2, \myvec{v}) \in \delta$,
\item $((q_1, post), a, (q_2, post), (\myvec{v}, 0))$ such that $(q_1, a, q_2, \myvec{v}) \in \delta$ and $a \in \Sigma$,
\item $((q_1, post), \separator, (q_2, post), (\myvec{v}, 0))$ such that $(q_1, \separator, q_2, \myvec{v}) \in \delta$.
\end{itemize}
From the construction of $\autb_{(p,q)}$, we can observe that the pair $(p,q)$ should satisfy that in $\aut$, there is a $\separator$-transition into $p$ and a $\separator$-transition out of $q$. Moreover, $q$ should be reachable from $p$ according to the  construction of $\aut_{R_2/R}[p, q] \cap \aut_1$. 

\begin{example}
Consider the constraint $\svarv = \mywrite(\svaru, 1, \svary)\ \wedge\ \svarv \in \separator (a^+ \separator)^*\ \wedge\ \mystrlen(\svarv) \ge 4$. Let $\aut = (\{r_1\}, \{q_0, q_1, q_2\}, \Sigma_{\separator}, \delta, \{q_0\}, \{q_1\})$ be the CEFA, where the register $r_1$ is used to record the string length and $\delta$ comprises the transitions $q_0 \xrightarrow[(1)]{\separator} q_1 \xrightarrow[(1)]{a}q_2 \xrightarrow[(1)]{a} q_2 \xrightarrow[(1)]{\separator} q_1$. 
Let us consider the pairs $(p, q)$ in $\aut$ satisfying that there is a $\separator$-transition into $p$ and a $\separator$-transition out of $q$, moreover, $q$ should be reachable from $p$.
One can easily observe that there is exactly one such pair, that is, $(q_1, q_2)$. 
Therefore, $\mywrite^{-1}(\Lang(\aut))$ comprises exactly one tuple $(\autb_{(q_1, q_2)}\cap \aut_0, \aut_{R_2/R}[q_1,q_2] \cap \aut_1, (r^{(1)}_1+r^{(2)}_1))$ such that
$\autb_{(q_1,q_2)} = (\{r^{(1)}_1, r' \}, Q', \Sigma_{\separator}, \delta', I', F')$ where 
\begin{itemize}
\item $Q' = \{q_0, q_1, q_2\} \times \{pre, idle, post\}$, $I' = \{(q_0, pre)\}$, $F' = \{(q_1, post)\}$, and 
\item $\delta' $ comprises the following transitions:
\begin{itemize}
\item $((q_1, pre), a, (q_2, pre), (1,0))$, $((q_2, pre), a, (q_2, pre), (1,0))$, 

\item $((q_0, pre), \separator, (q_1, pre), (1,1))$, $((q_2, pre), \separator, (q_1, pre), (1,1))$, 

\item $((q_0, pre), \separator, (q_1, idle), (1, 1))$ (since $(q_0, \separator, q_1, 1) \in \delta$),
\item $((q_1, idle), a', (q_1, idle), (0, 0))$ such that $a' \in \Sigma$, 
%
\item $((q_1, idle), \separator, (q_1, post), (1, 0))$  (since $(q_2, \separator, q_1, 1) \in \delta$),
\item $((q_1, post), a, (q_2, post), (1, 0))$ and $((q_2, post), a, (q_2, post), (1, 0))$,
\item $((q_2, post), \separator, (q_1, post), (1, 0))$.
%
\end{itemize}
\end{itemize}
To illustrate how $\autb_{(q_1, q_2)}$ and $\aut_{R_2/R}[q_1,q_2]$ work, we can see that $(\separator aa \separator a \separator, 6)$ is accepted by $\aut$, moreover,  for any $b\in\Sigma$,
\begin{itemize}[topsep=0pt,partopsep=0pt]
\item $(\separator ab \separator a \separator, 4, 1)$ is accepted by $\autb_{(q_1, q_2)}$, witnessed by 
\begin{center}
    $\begin{array}{l}
(q_0, pre) \xrightarrow[(1, 1)]{\separator} (q_1, idle) \xrightarrow[(0, 0)]{a} (q_1, idle) \xrightarrow[(0, 0)]{b} (q_1, idle) \\
\xrightarrow[(1, 0)]{\separator} (q_1, post) \xrightarrow[(1, 0)]{a} (q_2, post) \xrightarrow[(1, 0)]{\separator} (q_1, post),
\end{array}$
\end{center}
\item and $(aa, 2)$ is accepted by $\aut_{R_2/R}[q_1,q_2]$, witnessed by 
$
q_1 \xrightarrow[(1)]{a} q_2  \xrightarrow[(1)]{a} q_2
$.  
\end{itemize}
Thus, $\svaru$ and $\svary$  in $\svarv = \mywrite(\svaru, 1, \svary)$
can be $\separator ab \separator a \separator$ and $aa$, respectively.\qed

\end{example}


\paragraph*{Pre-image under $\subseq$.}
%
Let $\aut = (R, Q, \Sigma_{\separator}, \delta, I, F)$ be a CEFA. 
$\subseq^{-1}(\Lang(\aut))$ is computed as $(\autb\cap \aut_0, t)$ such that $t:= true$ and $\autb$ introduces two fresh registers $r'_1, r'_2$ to count the two numbers of occurrences of $\separator$ that correspond to the starting position and the length of the subsequence, respectively. Moreover, $\autb$ simulates the run of $\aut$ on the substring representing the subsequence returned by $\subseq$. 
Formally, $\subseq^{-1}(\Lang(\aut))$ is computed as $(\autb \cap \aut_0, true)$, where $\autb = (R \cup \{r'_1, r'_2\}, Q \cup \{pre, post\}, \Sigma_{\separator}, \delta', \{pre\}, \{post\})$ such that $pre, post$ are two fresh states denoting the fact that $\autb$ is reading a symbol before and after the subsequence respectively and $\delta'$ comprises the  following transitions:
\begin{itemize}[topsep=0pt,partopsep=0pt]
\item $(pre, a, pre, (\myvec{0}, 0,0))$ such that $a \in \Sigma$ (thus $a \neq \separator$),  
\item $(pre, \separator, pre, (\myvec{0}, 1,0))$,
\item $(pre, \separator, q, (\myvec{v}, 1,0))$ such that $(p, \separator, q, \myvec{v}) \in \delta$ for some $p \in I$, 
\item $(q, a, q', (\myvec{v}, 0, 0))$ such that $(q, a, q', \myvec{v}) \in \delta$ and $a \in \Sigma$, 
\item $(q, \separator, q', (\myvec{v}, 0,1))$ such that $(q, \separator, q', \myvec{v}) \in \delta$,
\item $(q, \separator, post, (\myvec{v}, 0,1))$ such that $(q, \separator, q', \myvec{v}) \in \delta$ for some $q' \in F$,
\item $(post, a, post, (\myvec{0}, 0,0))$ and $(post, \separator, post, (\myvec{0}, 0,0))$ where $a \in \Sigma$.
\end{itemize}
Note that in the backward propagation of $\aut$ with respect to an equality $\svarz = \subseq(\svarx, it_1, it_2)$, the constraint $r'_1 = it_1 \wedge r'_2 = it_2$ is added to the LIA constraint to assert that the value of $r'_1$ and $r'_2$ are equal to the beginning index and the length of the subsequence respectively.

\begin{example}
Let $\svarv = \subseq(\svaru, 2, 1) \wedge \svarv \in \separator (a^+ \separator)^* \wedge \mystrlen(\svarv) \ge 4$ and $\aut = (\{r_1\}, \{q_0, q_1, q_2\}, \Sigma_{\separator}, \delta, \{q_0\}, \{q_1\})$ be the CEFA, where the register $r_1$ is used to record the string length and $\delta$ comprises the transitions $q_0 \xrightarrow[(1)]{\separator} q_1 \xrightarrow[(1)]{a}q_2 \xrightarrow[(1)]{a} q_2 \xrightarrow[(1)]{\separator} q_1$. 
Then $\subseq^{-1}(\Lang(\aut))$ is $(\autb \cap \aut_0, true)$ such that 
\begin{center}
    $\autb = (\{r^{(1)}_1, r'_1, r'_2\}, \{q_0, q_1, q_2, pre, post\}, \Sigma_{\separator}, \delta', \{pre\}, \{post\})$
\end{center}
where 
$\delta' $ comprises the following transitions: 
\begin{itemize}[topsep=0pt,partopsep=0pt]
\item $(pre, a', pre, (0, 0,0))$ such that $a' \in \Sigma$, 
\item $(pre, \separator, pre, (0, 1,0))$,
\item $(pre, \separator, q_1, (1, 1,0))$ (since $(q_0, \separator, q_1, 1) \in \delta$), 
\item $(q_1, a, q_2, (1, 0, 0)$, $(q_2, a, q_2, (1, 0, 0))$, 
%
\item $(q_2, \separator, q_1, (1, 0, 1))$, 
%
\item $(q_2, \separator, post, (1, 0, 1))$ (since $(q_2, \separator, q_1, 1) \in \delta$ and $q_1$ is an accepting state in $\aut$), 
%
\item $(post, a', post, (0, 0,0))$ and $(post, \separator, post, (0, 0,0))$ where $a' \in \Sigma$.
\end{itemize}
To illustrate how $\autb$ works, we can see that $(\separator a  \separator a a \separator, 4, 2, 1)$ is accepted by $\autb$, witnessed by 
$$
pre \xrightarrow[(0, 1, 0)]{\separator} pre \xrightarrow[(0, 0, 0)]{a} pre \xrightarrow[(1, 1, 0)]{\separator} q_1 \xrightarrow[(1, 0, 0)]{a} q_2 \xrightarrow[(1, 0, 0)]{a} q_2  \xrightarrow[(1, 0, 1)]{\separator} post.  
$$
\end{example}



\paragraph*{Pre-image under $\elem$.}
Let $\aut = (R, Q, \Sigma, \delta, I, F)$ be a CEFA. 
$\elem^{-1}(\Lang(\aut))$ is constructed similar to $\subseq^{-1}(\Lang(\aut))$. Nevertheless, the construction is slightly different since the output of $\elem$ is a string, instead of a sequence. Intuitively, $\elem^{-1}(\Lang(\aut))$ is computed as $(\autb \cap \aut_0,t)$ where $t := true$, a fresh register $r'$ is introduced to count the number of occurrences of $\separator$ that corresponds to the position where the element is extracted. Formally, $\elem^{-1}(\Lang(\aut))$ is computed as $(\autb \cap \aut_0, true)$ such that $\autb = (R \cup \{r'\}, Q \cup \{pre, post\}, \Sigma_{\separator}, \delta', \{pre\}, \{post\})$, where $\delta'$ comprises the following transitions: 
\begin{itemize}[topsep=0pt,partopsep=0pt]
\item $(pre, a, pre, (\myvec{0}, 0))$ such that $a \in \Sigma$ (thus $a \neq \separator$),  
\item $(pre, \separator, pre, (\myvec{0}, 1))$,
\item $(pre, \separator, p, (\myvec{v}, 1))$ such that $p \in I$, 
%
\item $(q, a, q', (\myvec{v}, 0))$ such that $(q, a, q', \myvec{v}) \in \delta$ and $a \in \Sigma$ (thus $a \neq \separator$), 
\item $(q, \separator, post, (\myvec{0}, 0))$ such that $q \in F$,
\item $(post, a, post, (\myvec{0}, 0))$ and $(post, \separator, post, (\myvec{0}, 0))$ where $a \in \Sigma$.
\end{itemize}
Note that in the backward propagation of $\aut$ with respect to an equality $\svarz = \elem(\svarx, it)$, the constraint $r' = it$ is added to the LIA constraint to assert that the value of $r'$ is equal to the index $it$.

\hide{
\begin{example}
Consider the constraints $\svarv = \elem(\svaru, 2) \wedge \svarv = a$. Then the automaton $\aut$ that $\svarv$ belongs to is $q_0\xrightarrow{a}q_1$. The $\elem^{-1}(\aut)$ is $(\autb \cap \aut_0, true)$, where $\autb$ is $pre\xrightarrow[(1)]{\separator}pre\xrightarrow[(0)]{\Sigma}pre\xrightarrow[(1)]{\separator}q_0\xrightarrow[(0)]{a}q_1\xrightarrow[(0)]{\separator}post\xrightarrow[(0)]{\Sigma_{\separator}}post$. After back propagation, we remove the constraint $\svarv = \elem(\svaru, 1)$ and add the constraints $r' = 2\wedge \svaru \in \autb \cap \aut_0$, where $r'$ is the fresh register introduced in $\autb$. Solving the new constraints, we obtain a solution $\svaru = \separator\separator a \separator$ which represents the sequence $(\epsilon, a)$.
\end{example}
}

\medskip

Finally, we remark that the operation $\myseqlen(\svarx)$ can be modeled as a CEFA, similarly to the CEFA for $\mystrlen$ in \cite{atva2020}.

\section{Implementation and Experiments} \label{sect:exp}


We implement a solver  $\ostrichseq$ on top of the string solver OSTRICH~\cite{CHL+19} and the SMT solver $\princess$~\cite{princess08}, using about 5,000 lines of Scala code. The core functionality, preimage computation for dedicated string operations such as $\mywrite$, $\subseq$, $\elem$ and $\myseqlen$ is implemented in about 2,000 lines of code and is integrated directly into the OSTRICH framework. To support the representation of sequences, we built a lightweight library of automata which treat the separator as a special transition. The library is implemented in about 1,000 lines of code. 



\subsection{Benchmarks}

To evaluate the effectiveness of $\ostrichseq$, we curate two benchmark suites, $\seqbase$ and $\seqext$,
where $\seqbase$ only uses operations that are directly supported by some existing SMT solvers while $\seqext$ uses some string sequence operations that are not directly supported by any existing SMT solvers.

\smallskip
\noindent\emph{$\seqbase$.}
The $\seqbase$ benchmark suite comprises $\arraylogic$ formulas that contain only the generic sequence operations, that is, sequence operations that can be applied to the elements of any type (not necessarily string), such as sequence length $\myseqlen$, sequence concatenation $\cdot$, subsequence $seqt[it_1, it_2]$, sequence read $\myat(seqt, it)$ and sequence write $seqt[it \rightarrow strt]$.
$\seqbase$ is designed to facilitate a fair comparison with existing solvers since they do not support the string-specific sequence operations. 
$\seqbase$ contains 140 $\arraylogic$ constraints that are generated from three templates. 
\begin{itemize}
  \item  \polished{The first template $\svaru_1\in e_1 \wedge \svars_2 = \svars_1[n_1\rightarrow\svaru_1, n_2\rightarrow\svaru_1] \wedge \svaru_2 = \myat(\svars_2, m_1)\cdot\myat(\svars_2, m_2) \wedge \svaru_2 \in e_2 \wedge \myseqlen(\svars_2) < \mystrlen(\svaru_2)$ is utilized to curate 40 random  instances, where
  $\svaru_1$ is a string variable, $\svars_1,\svars_2$ are sequence variables, 
  and $e_1,e_2$ are randomly generated regular expressions.
  In the first 20 instances, $n_1, n_2, m_1, m_2$ are integer variables while in 
  the other 20 instances, they are random integers ranging from 0 to 10.}
        %
  \item \polished{The second template $\svaru_1\in e_1 \wedge \svars_2 = \svars_1[n_1\rightarrow\svaru_1, n_2\rightarrow\svaru_2] \wedge \svars_3 = \svars_2[n_3, len_1]\cdot\svars_2[n_4,len_2] \wedge \svaru_3 = \myat(\svars_3,m_1)\cdot\myat(\svars_3,m_2)\cdot\myat(\svars_3,m_3)\cdot\myat(\svars_3,m_4)\wedge\svaru_3\in e_2 \wedge \myseqlen(\svars_3) < \mystrlen(\svaru_3) + 1$ is utilized to curate 40 random instances, where $\svaru_1,\svaru_2$ are string variables, $\svars_1$, $\svars_2,\svars_3$ are sequence variables, and $e_1,e_2$ are randomly generated regular expressions.
  Similar to the first template,
  $n_i, m_i, len_1,len_2$ for $i\in [1,4]$  are integer variables in the first 20 instances but are random integers ranging from 0 to 10 in the other 20 instances.}
  \item \polished{The third template $\svars_1 = \svars_0[m_1\rightarrow \svaru_1]\wedge\cdots\wedge\svars_{i} = \svars_{i-1}[m_i\rightarrow \svaru_i]\wedge\svaru_1=\myat(\svars_1, n_1)\wedge\cdots\wedge\svaru_j=\myat(\svars_j, n_j)\wedge\svaru_1\in e_1 \wedge\cdots\wedge \svaru_j\in e_j$ is utilized to curate 60 random instances, where $i$ and $j$ are randomly selected from the range $0$ to $20$, $\svars_0 \cdots \svars_i$ are sequence variables,  $\svaru_1\cdots \svaru_j$ are string variables, $e_1\cdots e_j$ are randomly generated regular expressions, and 
  $m_1,\cdots m_i$, $n_1\cdots n_j$ are randomly generated integers constant ranging from 0 to 5. In this template, write ($\svars[\cdot\rightarrow\cdot]$) and read ($\myat(\svars,\cdot)$) operations are performed randomly on a string sequence, and the strings read from the sequence are checked against some regular expressions.}
\end{itemize}



\paragraph*{$\seqext$.}
The $\seqext$ benchmark suite comprises 60 $\arraylogic$ formulas that use specific string sequence operations, namely, $\myfilter_e$, $\mymatchall_e$, $\mysplit_e$ and $\myjoin_u$. Among them, 19 formulas are manually generated for the unit tests of $\ostrichseq$, and 41 formulas are generated from the real-world JavaScript programs collected from GitHub. Moreover, to compare with SMT solvers that do not support the considered string sequence operations, we rewrite them using the basic string operations whenever possible. ($\mymatchall_e$ is not rewritten since it cannot be expressed using existing string operations.) 

\subsection{Experimental results}

We conduct experiments by comparing $\ostrichseq$ with SOTA solvers, including cvc5, Z3, Z3-noodler, $\ostrich$ and $\princessarr$ (referring to an array-theory-based solver implemented within $\princess$). $\seco$~\cite{JezLMR23} is excluded 
due to soundness issues. Additionally, \verb|foldleft| and \verb|map| operations are not used to simulate $\myjoin$, as most constraints return unknown under such usage in Z3.
On $\seqbase$, we evaluate $\ostrichseq$ along with sequence solvers cvc5, Z3 and $\princessarr$. For $\seqext$, we rewrite the constraints using only the basic string operations and compare $\ostrichseq$ with string solvers cvc5, Z3, Z3-noodler and $\ostrich$.

All experiments are run on a 24 $\times$ 3.1 GHz-core server with 185 GB RAM, where each solver is started as a single thread, with a 60 seconds time limit and without the memory limit.

\begin{table}[t]
\setlength{\tabcolsep}{6pt}
  \centering
    \caption{\polished{Experimental results on $\seqbase$.}
  } 
    \begin{tabular}{| c | c | c | c | c |}
      \hline
                      & cvc5        & Z3    & $\princessarr$ & $\ostrichseq$ \\
      \hline
      sat             & 48          & 14    & \textbf{52}    & \textbf{52}   \\
      \hline
      unsat           & 87          & 59    & 77             & \textbf{88}   \\
      \hline
      solved          & 135         & 73    & 129            & \textbf{140}  \\
      \hline
      unknown/timeout & 5           & 67    & 11             & \textbf{0}    \\
      \hline
      avg. time (s)   & 3.53        & 24.36 & 6.84           & \textbf{3.23}          \\
      \hline
    \end{tabular} %
  \label{tab:seq-base}
\end{table}

The results on $\seqbase$ are reported in Table~\ref{tab:seq-base}.
We can see that $\ostrichseq$ outperforms all the other solvers on $\seqbase$ in terms of both the number of solved instances and the solving time per instance. Specifically,   $\ostrichseq$ solves all the \polished{140} instances in $\seqbase$, while cvc5, Z3 and $\princessarr$ only solve \polished{135}, \polished{73} and \polished{129} instances, respectively. Moreover,  $\ostrichseq$ is more efficient than cvc5, Z3 and $\princessarr$ ($\ostrichseq$ spends \polished{3.23} seconds in solving each instance from $\seqbase$ on average, while cvc5, Z3 and $\princessarr$ spend \polished{3.25}, \polished{24.36}, and \polished{6.84} seconds, respectively). 


The results on $\seqext$ are reported in Table~\ref{tab:seq-ext}. Since the SOTA sequence solvers do not support $\myfilter_e$, $\mymatchall_e$ or $\mysplit_e$ directly, the $\arraylogic$ constraints are rewritten to string constraints (instead of solving the $\arraylogic$ constraints directly) when we compare with them on $\seqext$.
Moreover, while we can use {\sf str.replace\_re\_all} to encode $\myfilter_e$ and $\mysplit_e$, the encoding of $\mymatchall_e$ requires finite-state transducers which have not been supported by most SOTA string solvers. 
As a result, we do \emph{not} generate the string constraints for the 12 instances in $\seqext$ that contain $\mymatchall_e$, which are denoted by 
$\seqext\mbox{\sf -M}$, whereas the rest 48 instances are 
denoted by $\seqext\mbox{\sf -N}$. 

\begin{table}[t]
\setlength{\tabcolsep}{2pt}
\caption{Experimental results on $\seqext$.
}
\centering
    \begin{tabular}{| c| c| c | c | c | c| c|}
      \hline
                                              &                 & cvc5  & Z3   & Z3-noodler    & $\ostrich$ & $\ostrichseq$ \\
      \hline
      \multirow{3}{*}{$\seqext\mbox{\sf -N}$} & sat             & 19    & 4    & 4             & 20         & \textbf{23}   \\
      \cline{2-7}
                                              & unsat           & 21    & 6    & 4             & 18         & \textbf{25}   \\
      \cline{2-7}
                                              & solved          & 40    & 10   & 8             & 38         & \textbf{48}   \\
      \cline{2-7}
                                              & unknown/timeout & 8     & 38   & 40            & 10         & \textbf{0}    \\
      \cline{2-7}
                                              & avg. time (s)   & 10.83 & 0.13 & \textbf{0.05} & 15.92      & 3.53          \\
      \hline
      \multirow{3}{*}{$\seqext\mbox{\sf -M}$} & sat             & -     & -    & -             & -          & \textbf{12}   \\
      \cline{2-7}
                                              & unsat           & -     & -    & -             & -          & \textbf{0}    \\
      \cline{2-7}
                                              & solved          & -     & -    & -             & -          & \textbf{0}    \\
      \cline{2-7}
                                              & unknown/timeout & -     & -    & -             & -          & \textbf{0}    \\
      \cline{2-7}
                                              & avg. time (s)   & -     & -    & -             & -          & \textbf{1.77} \\
      \hline
    \end{tabular}
  \label{tab:seq-ext}
\end{table}

From Table~\ref{tab:seq-ext}, we can see that Z3 and Z3-noodler solve only 10 and 8 out of 48 constraints in $\seqext\mbox{\sf -N}$, respectively, and cvc5 and $\ostrich$ solve 40 and 38 instances, respectively,
indicating that the complexity of the string constraints generated from the $\arraylogic$ constraints. 
On the other hand, $\ostrichseq$ solves all of them, demonstrating its superiority in solving the $\arraylogic$ constraints with string sequence operations. For efficiency, 
$\ostrichseq$ spends 3.53 seconds per instance on average, significantly less than cvc5 and $\ostrich$ (10.83 and 15.92 seconds on average, respectively). For the constraints in $\seqext\mbox{\sf -M}$, $\ostrichseq$ successfully solves all 12 instances with an average time of 1.77 seconds.
As aforementioned, for the constraints in $\seqext\mbox{\sf -M}$, we do not generate string constraints or compare against cvc5, Z3, Z3-noodler and $\ostrich$, as they cannot be expressed using only the basic operations of string theory.




\section{Related work}\label{sect:related}


{\bf String constraint solving.} 
String constraint solving has received considerable attention in the literature. Amadini \cite{Amadini23} provides a comprehensive survey of the literature up to 2021. In particular, 
prior approaches 
are classified into three main categories: automata-based (relying on finite automata to represent the domain of string variables
and to handle string operations, e.g., \cite{Abdulla14,HJLRV18,CCH+18,Vijay-length,LB16,BerzishKMMDNG20,ChenCHHLS24}), word-based (algebraic approaches based on systems of word equations, e.g., \cite{cvc4,BarbosaBBKLMMMN22,Z3,Z3str2,Z3str3,Z3str4,TCJ16,S3,BerzishKMMDNG20,ChenCHHLS24}), 
and unfolding-based approaches (explicitly reducing each string into a number of contiguous elements denoting its
characters, e.g., \cite{SFPS17,KGAGHE13,SAHMMS10,LG13,AGS18,DEKMNP19}). 

From another perspective, there are two lines of work aiming for building practical string constraint solvers. 
(1) One could support as many operations as possible, but primarily resort to heuristics, offering no completeness/termination guarantees. 
Many string constraint solvers are implemented in SMT solvers, 
allowing combination with other theories, most commonly the theory of integers for string lengths. Some (non-exhaustive) examples include cvc4/5 \cite{cvc4,BarbosaBBKLMMMN22}, Z3 \cite{Z3,BTV09}, Z3-str2/3/4 \cite{Z3str2,Z3str3,Z3str4}, S3(P)~\cite{TCJ16,S3}, Trau~\cite{Abdulla17} (or its variants Trau+ \cite{AbdullaA+19}), 
Slent~\cite{WC+18}, etc.  
%
%
More recent ones include Z3str3RE \cite{BerzishKMMDNG20} 
%
and Z3-Noodler \cite{ChenCHHLS24} 
(which is based on stabilization-based algorithm \cite{BlahoudekCCHHLS23,ChenCHHLS23} for solving word equations with regular constraints). 
%
%
(2) The second approach is to develop solvers for decidable fragments, 
including Norn \cite{Abdulla14}, SLOTH \cite{HJLRV18} and
OSTRICH \cite{CCH+18,CHL+19}, which are usually based on complete decision procedures (e.g. \cite{Vijay-length,AbdullaA+19,LB16}) 


We also mention that there are solvers which emphasize certain aspects of string constraint solving by providing dedicated methods or optimization. These include, for instance, those for more expressive regular expressions \cite{BerzishDGKMMN23,ChenFHHHKLRW22},  regex-counting~\cite{HuW23},  
string/integer conversion and flat regular constraints~\cite{WuCWXZ24}, Not-Substring Constraint~\cite{AbdullaACDHHTWY21}, integer data type~\cite{atva2020}, etc. 


\smallskip
\noindent
{\bf Sequences as an extension of strings.} Strings and sequences are closely related, but may exhibit in different forms. For instance, a string can be considered as a sequence over a finite alphabet. 
Jez et al.~\cite{JezLMR23} recently studied sequence theories which are an extension of theories of strings with an infinite alphabet of letters, together with a corresponding alphabet theory (e.g. linear integer arithmetic). They provided decision procedures based on parametric automata/transducers, giving rise to a sequence constraint solver $\seco$. Our work is different in that we essentially work on a sequence theory where the element is instantiated by the string type.

\smallskip
\noindent
{\bf Theory of arrays.}
There are many studies on solving formulas in the theory of arrays or its extensions, for instance, \cite{SBDL01,BMS06,HIV08,FalkeMS13,DacaHK16,FarinierDBL18}, which are also related to the sequence theory. Note that in these studies, the elements are considered to be either of a generic type or of integer type. In other words, the theory of arrays where the elements are of the string type has not been defined and investigated therein. 

\section{Conclusion} \label{sect:conc}

In this paper, we have proposed $\arraylogic$, a logic of string sequences, which supports a wealth of string and sequence operations---especially their interactions---commonly in real-word string-manipulation programs. We have provided a decision procedure for the straight-line fragment, which is implemented as a new tool $\ostrichseq$. The experiments on both hand-crafted and real-world benchmarks demonstrate that $\ostrichseq$ is an effective and promising tool for testing, analysis and verification of string-manipulation programs in practice. 

%
\newpage

\appendix
\section{Appendix: Representing $\myfilter_e$, $\mysplitstr_e$, $\mymatchallstr_e$ and $\myjoin_u$ by NFT} \label{sect:appdenix}
The semantics of regular expression matching, like first match and longest match, can be represented by transducers (e.g., refer to~\cite{min2010transducer}). While any matching semantics could be used, we specifically illustrate the construction of transducers utilizing left and longest match semantics. The construction detailed in this section also utilizes $\epsilon$-transitions in (nondeterministic) finite transducers (NFTs), which does not exceed the expressive capabilities of the NFTs described in Section \ref{sec:prelim}.
\subsection{Representing $\myfilter_e$ by NFT}
Recall that $\myfilter_e(s)$ removes every substring of $s$ between two consecutive occurrences of $\separator$ that does not match the regular expression $e$. To build the NFT that represents $\myfilter_e$, we assume that the automata $\aut = (Q,\Sigma,\delta,I,F)$ and $\Bar{\aut}=(\Bar{Q},\Sigma,\Bar{\delta},\Bar{I},\Bar{F})$ are already built to identify the language of $e$ and $\Bar{e}$, respectively. We then introduce four states: $q_{n}$ for the start point of searching on $s$, $q_{b}$ for beginning to match $e$, $q_{f}$ for a failed match of $e$, and $q_t$ for the terminal of the searching. The core idea is to nondeterministically select either $e$ or $\Bar{e}$ for a substring between two $\separator$ symbols and output the substring if it matches $e$ successfully; otherwise output an empty string. Formally, the NFT illustrating $\myfilter_e$ is defined as $(\{q_{n}, q_{b}, q_{f}, q_{t}\}\cup Q \cup \Bar{Q}, \Sigma_{\separator}, \delta', \{q_{n}\}, \{q_{t}\})$ where $\delta$ comprises  tuples that correspond to the matching of $e$
\begin{itemize}
    \item $(q_n,\separator, q_b, \separator)$,
    \item $(q_b, \epsilon, q_I, \epsilon)$ such that $q_I\in I$,
    \item $(q, a, q', a)$ such that $(q,a,q') \in \delta$,
    \item $(q_F, \separator, q_b, \separator)$ such that $q_F\in F$,
    \item $(\Bar{q}_F, \separator, q_b, \separator)$ such that $\Bar{q}_F \in \Bar{F}$,
\end{itemize}
and tuples that indicate a mismatch of $e$
\begin{itemize}
    \item $(q_n,\separator, q_f, \epsilon)$,
    \item $(q_f, \epsilon, \Bar{q}_I, \epsilon)$ such that $\Bar{q}_I\in \Bar{I}$,
    \item $(\Bar{q}, a, \Bar{q'}, \epsilon)$ such that $(\Bar{q}, a, \Bar{q'})\in \Bar{\delta}$,
    \item $(q_F, \separator, q_f, \epsilon)$ such that $q_F \in F$,
    \item $(\Bar{q}_F, \separator, q_f, \epsilon)$ such that $\Bar{q}_F \in \Bar{F}$,
\end{itemize}
and tuples that terminate the search on $s$
\begin{itemize}
    \item $(q_n,\separator, q_t, \separator)$,
    \item $(q_F,\separator, q_t, \separator)$ such that $q_F \in F$,
    \item $(\Bar{q}_F,\separator, q_t, \separator)$ such that $\Bar{q}_F \in \Bar{F}$.
\end{itemize}

\subsection{Representing $\mymatchallstr_e$ by NFT}
Recall that $\mymatchallstr_e(u)$ transforms $u = u'_1 v_1 u'_2 v_2 \cdots u'_{k-1}v_{k-1} u'_k$ into $\separator v_1$ $ \separator \cdots \separator v_{k-1} \separator$, where for each $i \in [k-1]$, $v_i$ is the leftmost and longest occurrence of $e$ in \polished{$u'_i v_i u'_{i+1}\cdots v_{k-1}u'_k$}, moreover, $u'_k$ does not contain any substring that matches $e$. Here, we must preserve the leftmost and longest match semantics. Therefore, we employ three modes to show the matching situations: $\nmatch$ to indicate no match for $e$, $\match$ to signify that $e$ is currently matched, and $\ematch$ to denote the completion of matching $e$. Assume that $\aut = (Q,\Sigma,\delta,I,F)$ is the automaton that recognizes $e$, $Q' = \{\langle mode, cur, noreach\rangle$ $\mid mode\in \{$\nmatch$, $\match$, $\ematch$\}, cur\in 2^Q, noreach \in 2^Q\}$ is the primary states set in the NFT representing $\mymatchallstr_e$. The intuition is utilizing $cur$ to denote the states matched currently and $noreach$ to denote the states unable to reach final states, thereby ensuring
\begin{itemize}
    \item whenever the NFT transitions from mode $\nmatch$ to $\match$, any current states must not reach final states (otherwise we do not match leftmost),
    \item whenever the NFT transitions from mode $\match$ to $\ematch$, any current states must not reach final states again (otherwise we do not match longest).
\end{itemize}
Let $next_a(Q_0) = \{q' \mid (q, a, q') \in \delta, q \in Q_0\}$ denote the set of states reachable from $Q_0$ after reading label $a$ within $\aut$, then the NFT illustrating $\mymatchallstr_e$ is $(Q'\cup \{q_f\}, \Sigma_\separator, \delta', \langle\nmatch,\emptyset,\emptyset\rangle, \{q_f\})$ where $F'$ comprises states
$\delta'$ comprises tuples that indicate the termination of the search
\begin{itemize}
    \item $(\langle\langle\nmatch, \emptyset, noreach\rangle, \epsilon, q_f, \separator)$ such that $noreach\cap F=\emptyset$,
    \item $(\langle\ematch, cur, noreach\rangle, \epsilon, q_f, \separator)$ such that $noreach\cap F=\emptyset$ and $cur \in 2^Q$,
\end{itemize}
and tuples illustrating the transformation among these three modes
\begin{itemize}
    \item $(\langle\nmatch, \emptyset, noreach\rangle, a, \langle\nmatch, \emptyset, next_a(noreach\cup I)\rangle, \epsilon)$,
    
    \item $(\langle\nmatch, \emptyset, noreach\rangle, a, \langle\match, next_a(I), next_a(noreach)\rangle, \separator a)$ \\ such that $next_a(I)\not=\emptyset$,
    \item $(\langle\nmatch, \emptyset, noreach\rangle, a, \langle\ematch, next_a(I), next_a(noreach)\rangle, \separator a)$ \\ such that $next_a(I)\cap F\not= \emptyset$,
    
    \item $(\langle\match, cur, noreach\rangle, a, \langle\match, next_a(cur), next_a(noreach)\rangle, a)$ \\ such that $next_a(cur)\not=\emptyset$,
    \item $\langle(\match, cur, noreach\rangle, a, \langle\ematch, next_a(cur), next_a(noreach)\rangle, a)$ \\ such that $next_a(cur)\cap F\not= \emptyset$,
    
    \item $(\langle\ematch, cur, noreach\rangle, a, \langle\nmatch, \emptyset, next_a(cur\cup noreach\cup I)\rangle, \epsilon)$, 
    
    \item $(\langle\ematch, cur, noreach\rangle, a, \langle\match, next_a(I), next_a(cur\cup noreach)\rangle, \separator a)$ \\ such that $next_a(I)\not=\emptyset$,
    \item $\langle(\ematch, cur, noreach\rangle, a, \langle\ematch, next_a(I), next_a(cur\cup noreach)\rangle, \separator a)$ \\ such that $next_a(I)\cap F\not= \emptyset$.
    
\end{itemize}

\subsection{Representing $\mysplitstr_e$ and $\myjoin_u$ by NFT}
$\mysplitstr_e(u)$ transforms $u = u'_1 v_1 u'_2 \cdots u'_{k-1} v_{k-1} u'_k$ into  $\separator u'_1 \separator u'_2 \cdots u'_{k-1} \separator u'_k \separator$, where for each $i \in [k-1]$,  \polished{$v_i$} is the leftmost and longest matching of $e$ in \polished{$u'_i v_i u'_{i+1}\cdots v_{k-1}u'_k$}. $\mysplitstr_e(u)$ can be simulated by $\separator\cdot \mathsf{replaceAll}_e(u, \separator)\cdot\separator$. Here, $\mathsf{replaceAll}_e(u, \separator)$ denotes the function replacing each occurrence of $e$ in $u$ with $\separator$, which can be seen as a transducer \cite{CCH+18}.

$\myjoin_u(v)$ removes the first and the last occurrences of $\separator$ and replaces all the remaining occurrences of $\separator$ in $v$ by $u$. Similar to $\mysplitstr_e$, $\myjoin_u(v)$ can be emulated using $\mathsf{replaceAll}$ function along with concatenation, e.g., $v' = \myjoin_u(v)$ can be transformed to $v = \separator\cdot v_{tmp} \cdot \separator\wedge v' = \mathsf{replaceAll}_\separator(v_{tmp}, u)$.

Considering the observation mentioned above, the NFTs that represent $\mysplitstr_e$ and $\myjoin_u$ are straightforward.

\bibliographystyle{abbrv}
\bibliography{APLAS2025/ref}

\end{document}